\documentclass[doublecol]{epl2} 
\usepackage{amsmath}
\usepackage{graphicx}
\usepackage{amsfonts}
\usepackage{amssymb}
\usepackage{amsmath}
\usepackage{bbold}
\usepackage{color}
\usepackage{comment}

\newcommand{\up}{\uparrow}
\newcommand{\down}{\downarrow}
\renewcommand{\k}{{\bf k}}
\newcommand{\eb}{\varepsilon_B}
\newcommand{\p}{{\bf p}}
\newcommand{\q}{{\bf q}}
\newcommand{\0}{{\bf 0}}

\newcommand{\R}{\bf R}
\renewcommand{\r}{{\bf r}}

\newcommand{\eq}{\epsilon_{\q\up}}

\newcommand{\ep}{\epsilon_{\p\up}}
\newcommand{\ek}{\epsilon_{\k\up}}

\newcommand{\T}{{\cal T}}

\title{Three-body problem in a two-dimensional Fermi gas}

\author{Vudtiwat Ngampruetikorn\inst{1} \and Meera M. Parish\inst{1,2}  \and Jesper Levinsen\inst{1}}
\shortauthor{V. Ngampruetikorn, M. M. Parish and J. Levinsen}

\institute{                    
  \inst{1} T.C.M. Group, Cavendish Laboratory, JJ Thomson Avenue, Cambridge,
 CB3 0HE, United Kingdom\\
  \inst{2} London Centre for Nanotechnology, Gordon Street, London, WC1H 0AH, United Kingdom
}

\pacs{34.50.-s}{Atomic and molecular scattering}
\pacs{31.15.ac}{Few-body systems}
\pacs{05.30.Fk}{Fermion systems (quantum statistical mechanics)}

\date{\today}

\abstract{
  We investigate the three-body properties of two identical $\uparrow$
  fermions and one distinguishable $\downarrow$ atom interacting in a
  strongly confined two-dimensional geometry.  We compute exactly the
  atom-dimer scattering properties and the three-body recombination
  rate as a function of collision energy and mass ratio
  $m_\uparrow/m_\downarrow$.  We find that the recombination
    rate for fermions is strongly energy dependent, with significant
    contributions from higher partial waves at low energies. 
For $m_\up\lesssim m_\down$, the $s$-wave atom-dimer scattering below
  threshold is completely described by the scattering length.
  Furthermore, we examine the $\uparrow\uparrow\downarrow$ bound
  states (trimers) appearing at large $m_\uparrow/m_\downarrow$ and
  find that the energy spectrum for the deepest bound trimers
  resembles that of a hydrogen atom confined to two dimensions.
}

\begin{document}

\maketitle

\section{Introduction}
Ultracold atomic gases have proven to be an extremely
versatile system, providing elegant realizations of a range of quantum
many-body phenomena such as the BCS-BEC crossover and the Mott
transition~\cite{RevModPhys.80.1215,RevModPhys.80.885}.  In
particular, the cold-atom system involves short-range interactions
that can be tuned to have large scattering lengths, thus rendering the
low-energy physics insensitive to the details of the interaction
potentials and essentially ``universal''. This has enabled the study
of universal few-body physics, which in turn has had major
consequences for the many-body system. For instance, the scattering
length of diatomic molecules (dimers) was necessary for a complete
description of the BCS-BEC
crossover~\cite{PetrovPRL2004,brodsky2006,levinsen2006}.

Few-body inelastic processes in the cold-atom system are also
important since they limit the lifetime of the gas and constrain the
densities that can be achieved in experiment.
Furthermore, they can act as an indirect probe of the quantum system,
e.g., the first experimental evidence for the Efimov
effect~\cite{1973NuPhA.210..157E} was deduced from three-body
losses~\cite{2006Natur.440..315K}.
There is even the prospect of generating strongly correlated phases
using few-body loss processes: dimer-dimer dissipation has already
been shown experimentally to induce correlations~\cite{Syassen2008},
while it has recently been proposed that three-body dissipation can be
used to engineer the Pfaffian state in two dimensions
(2D)~\cite{cirac2009}.

In this paper, we  investigate the universal three-body problem of two identical $\uparrow$ fermions and 
one $\downarrow$ particle in 2D.
Such an investigation is timely and important given the recent experiments on Fermi gases
confined to a quasi-2D geometry~\cite{PhysRevLett.105.030404,PhysRevLett.106.105301,
PhysRevLett.106.105304,2011Natur.480...75F,
  PhysRevLett.108.045302,2012Natur.485..619K,PhysRevLett.108.235302}.
Moreover, the few-body properties in 2D are qualitatively different from those in 3D:
e.g., for identical bosons, 
the rate of recombination of three atoms into an atom and a dimer
vanishes at low energies~\cite{PhysRevA.83.052703}, in contrast to the 3D case.
Indeed, we show here that the recombination rate for fermions 
has a similar dependence on energy to the case of identical
bosons, 
{but surprisingly we also find that higher partial wave
  contributions can be significant at low energies for both bosons and
  fermions, while the fermionic rate can even be larger than
  that for identical bosons in this regime.}  This result could
potentially be important for any 2D experiment seeking to realize
itinerant ferromagnetism~\cite{2011NJPh...13e5003Z}.  Furthermore,
bound states can impact the scattering properties; in particular,
the presence of $\uparrow\uparrow\downarrow$ bound states (non-Efimov
trimers) for mass ratios $m_\uparrow/m_\downarrow \gtrsim
3.33$~\cite{PhysRevA.82.033625} is shown to lead to an enhanced
$p$-wave scattering cross section in the scattering of a $^{40}$K atom
and a $^{40}$K-$^6$Li dimer.  We additionally find that the deepest
bound trimers have an energy spectrum that resembles that of a
hydrogen atom confined to 2D, unlike the case in 3D where Efimov
physics dominates.


\section{Model}
In the following, we consider a two-component Fermi gas confined to 
2D by a strong, approximately harmonic confinement,
$V_{\up,\down}(z)=\frac12m_{\up,\down}\omega_{\up,\down}^2z^2$. The
gas can be considered to be kinematically 2D when the temperature and
Fermi energy are both smaller than the confining frequencies,
$\omega_{\up,\down}$ (we set $k_\text{B}=\hbar=1$).  The $\up$-$\down$
interaction is characterized by a 3D $s$-wave scattering length $a_s$
much larger than the van der Waals range of interatomic forces.  The
two species $\up,\down$ are either different hyperfine states of
the same atom, or single hyperfine states of different atomic species
such as $^6$Li and $^{40}$K.  The low-energy scattering of two atoms
is described through the $s$-wave scattering amplitude
\begin{equation}
f_{\uparrow\downarrow}(q)=\frac{2\pi}{\ln\left[1/(qa_{2\text{D}})\right]+i\pi/2},
\label{eq:f}
\end{equation}
where $\q$ is the relative momentum. 
We see here that the interparticle
interaction is fundamentally different from the 3D case: it is
energy dependent even at low energies and 
there always exists a bound state with binding energy $\varepsilon_B=1/(2\mu a_{2\text{D}}^2)$,
where the reduced mass $\mu^{-1}=m_\up^{-1}+m_\down^{-1}$. 
Equivalently, we can
define the scattering $T$-matrix which
describes the repeated interparticle scattering at total energy $E$
and momentum~$\q$, 
\begin{equation} \notag
  \T(\q,E)=\frac{2\pi/\mu}{-\ln\left[(E-q^2/2(m_\up+m_\down)+i0)/\eb\right]+i\pi}.
\end{equation}
Assuming that the confinement frequencies are identical for the two
species, then in the limit
$\eb\ll\omega_z$, the parameter $a_{2\text{D}}$ is related to the 3D
scattering length by $a_{2\text{D}}=l_z
\sqrt{\pi/B}\exp(-\sqrt{\pi/2}l_z/a_s)$, with 
$B \approx 0.905$~\cite{PhysRevA.64.012706,RevModPhys.80.885} and confinement length
$l_z=1/\sqrt{2\mu\omega_z}$.
For simplicity, in this work we restrict ourselves to the 2D limit. However,
deviations from the 2D limit may be captured by the use of a
two-channel model where $l_z$ plays the role of an effective range in
the two-body $T$-matrix \cite{Levinsen:2012fk}.

\begin{figure}
\centering
\includegraphics[width=\linewidth]{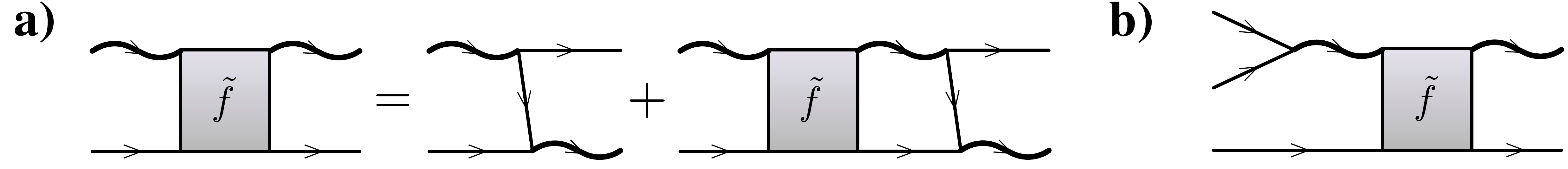}
\caption{
(a) Illustration of the STM 
 equation.
Wavy (straight) lines are the $T$-matrix (atom propagator),
respectively. (b) The three-body recombination process.
\label{fig:feynman}}
\end{figure}


\section{Three-body problem}
We now consider the scattering of a spin-$\up$ atom with an $\up\down$
dimer. The process is illustrated in Fig.~\ref{fig:feynman}(a) and is
described by the Skorniakov--Ter-Martirosian (STM) integral equation~\cite{stm}. 
We let the incoming [outgoing] atom and dimer have four-momenta
$(\k,\ek)$ and $(-\k,E-\ek)$ [$(\p,\ep)$ and $(-\p,E-\ep)$],
respectively, with total energy $E = k^2/2\mu_3 - \varepsilon_B$ to
ensure that the dimer is on-shell.  Here we define
$\epsilon_{\k\up,\down}=k^2/2m_{\up,\down}$ and the atom-dimer reduced
mass $\mu_3^{-1} = (m_\uparrow+m_\downarrow)^{-1} + m_\uparrow^{-1}$.
With these definitions, the STM equation in 2D reads
\begin{equation}
\tilde f_\ell(k,p) = \zeta h(k,p)\Big[
 g_\ell(k,p)
  - \int  \frac{q\, d q}{2\pi} \,
\frac{g_\ell(p,q)\tilde f_\ell(k,q)}{q^2-k^2-i0}\Big].
\label{eq:stm}
\end{equation}
The atom-dimer scattering preserves angular momentum and we have used
this to decouple the scattering amplitude $\tilde f$ into partial
waves, denoted by $\ell=0$ for $s$-wave, $\ell=1$ for $p$-wave, etc.
The projection of the spin-$\down$ atom propagator onto the $\ell$'th
partial wave is given by
\begin{equation}
g_\ell(p,q)=\int_0^{2\pi}\frac{\cos(\ell\phi)d\phi/2\pi}{
E-\ep-\eq-\epsilon_{\p+\q\down}+i0},
\end{equation}
with $\phi$ the angle between $\p$ and $\q$. We define the function
$h(k,p)\equiv (k^2-p^2)\T(\p,E-\ep) $ to separate out the simple pole
of the two-particle propagator occuring at $|\k|=|\p|$.  
We also include the factor $\zeta$ describing the quantum
statistics. It takes the value $-1$ when the spin-$\up$ particles are
fermions, $+2$ for three identical bosons, and $+1$ in the case of
heteronuclear bosons with negligible $\up$-$\up$ interaction.

\begin{figure}[t]
\centering
\includegraphics[width=.9\linewidth]{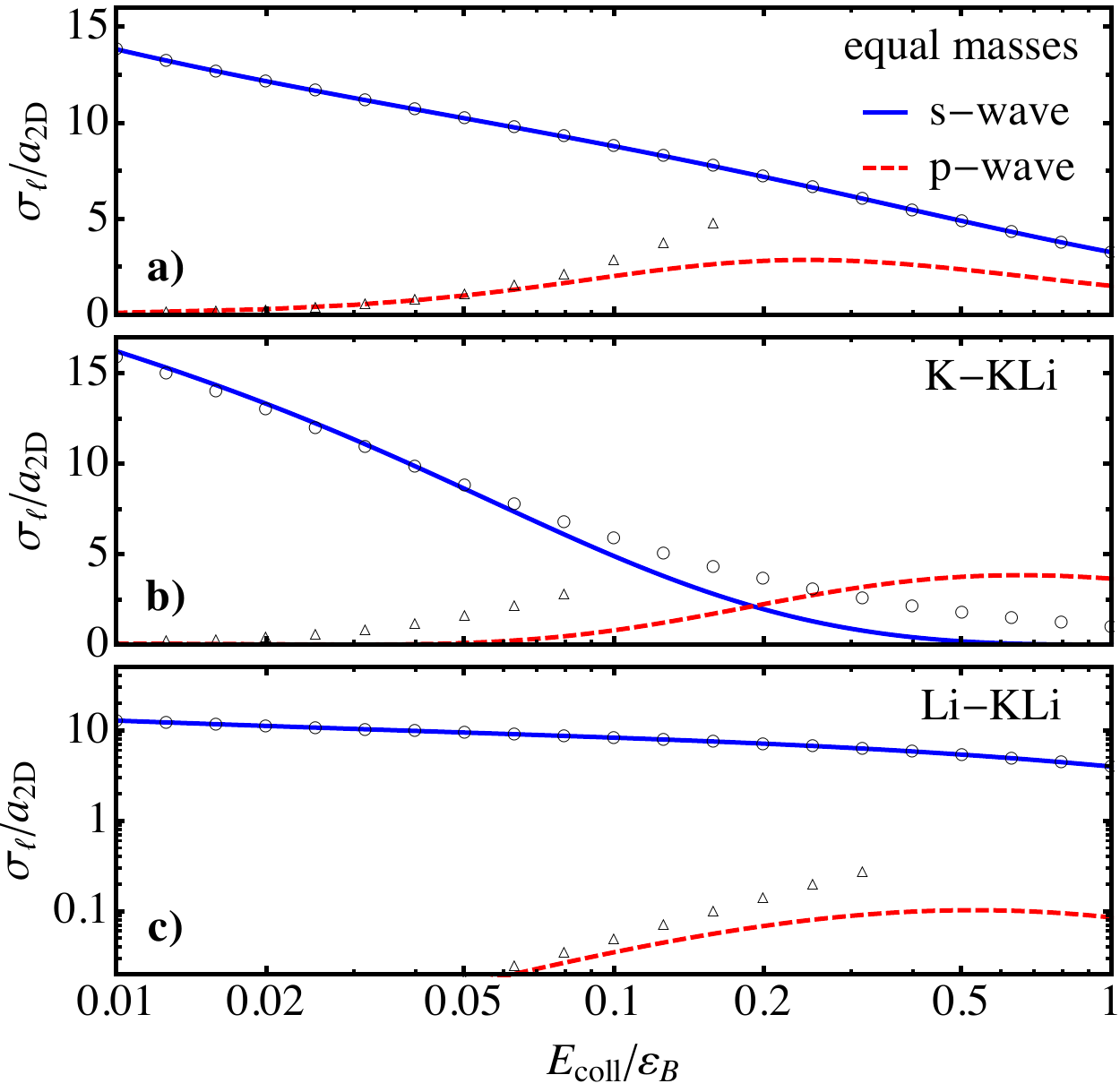}
\caption{(Color online) $s$- and $p$-wave elastic scattering cross
  sections
for three different mass
  ratios. Circles and triangles correspond to the low-energy asymptotic behavior,
  Eq.~(\ref{eq:asymp}). Note the log-scale in (c).
\label{fig:elastic}}
\end{figure}
%
 

\section{Elastic atom-dimer scattering}
The elastic scattering cross section in partial wave $\ell$
takes the form~\cite{LL3:1977,1986AmJPh..54..362A,note1}
\begin{equation}
\sigma^\text{el}_\ell(E) = \frac{|f_\ell(k)|^2}{4k}(2-\delta_{\ell,0}).
\end{equation}
Note that this has dimensions of length in 2D.
The partial wave scattering amplitudes are related to the
solutions of Eq.~(\ref{eq:stm}) by $f_\ell(k) =
\tilde f_\ell(k,k)$. 
Figure~\ref{fig:elastic} shows the $s$- and $p$-wave scattering cross
sections as a function of collision energy, $E_\text{coll}
=k^2/2\mu_3$, for equal masses and for the mass ratios corresponding
to a $^6$Li-$^{40}$K heteronuclear Fermi mixture.  We see that
$s$-wave scattering always dominates when $m_\up\leq m_\down$,  
while $p$-wave scattering becomes dominant for the K-KLi atom-dimer
scattering when $E_\text{coll}/\eb\geq 0.2$. 
This is related to a 
$p$-wave trimer state crossing the atom-dimer continuum 
at the mass ratio 3.33, 
leading to a resonant enhancement of $p$-wave scattering. A similar
enhancement of $p$-wave scattering was predicted in
3D~\cite{PhysRevLett.103.153202} for the K-KLi system due to the
appearance of a trimer at a mass ratio of
8.2~\cite{2007JPhB...40.1429K}. 
The presence of trimers in higher odd partial waves in 2D
at large mass ratio~\cite{PhysRevA.82.033625} will likewise lead to a
resonantly enhanced atom-dimer interaction.

\begin{figure}[t]
\centering
\includegraphics[width=.95\linewidth]{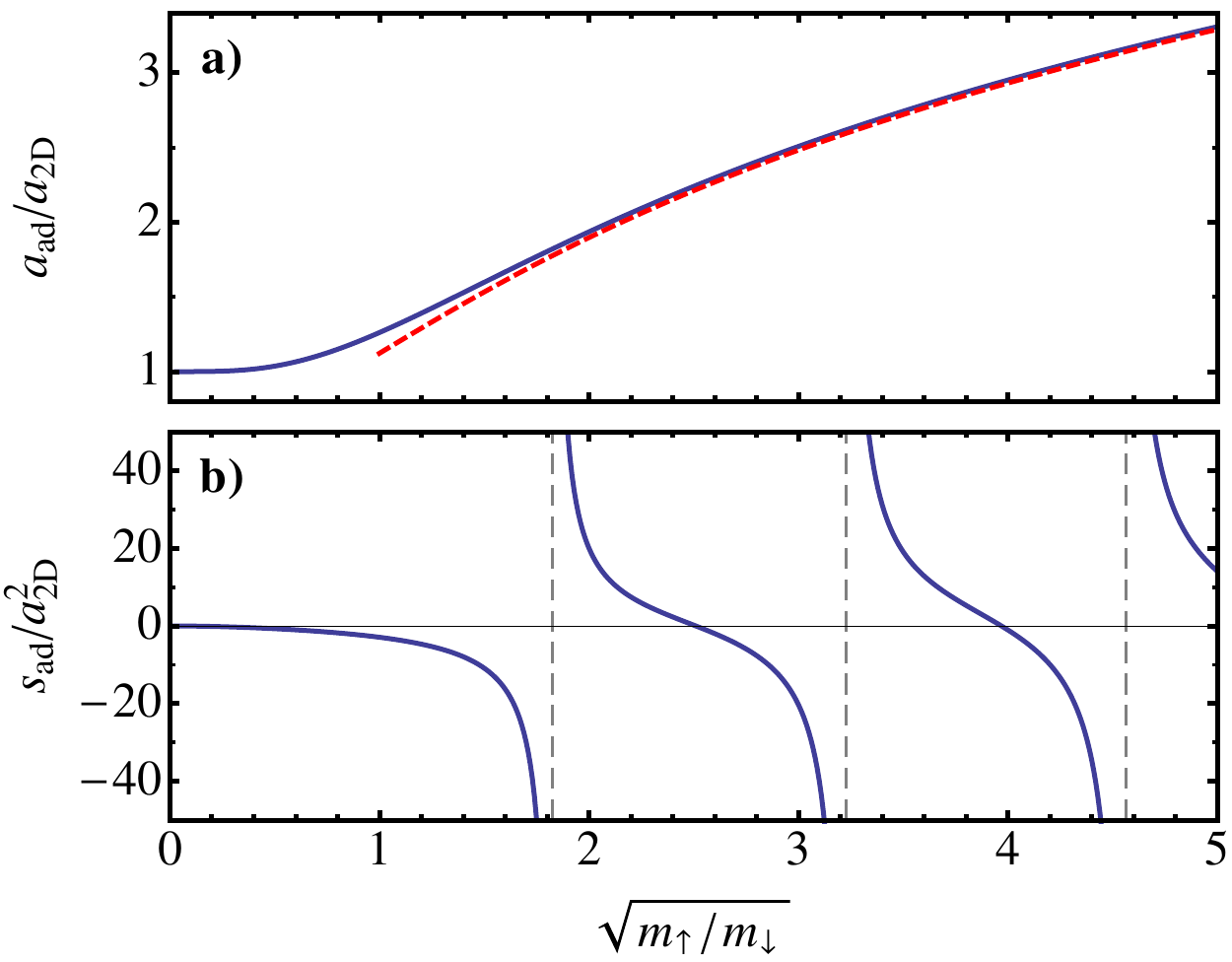}
\caption{(Color online) (a) Scattering length and (b) surface as a
  function of mass ratio. The dashed line in (a) is calculated from the asymptotic
  expression in the limit of a large mass imbalance. The vertical dashed lines
  in (b) signify the appearance of trimers.
  \label{fig:lowe}}
\end{figure}

In the low-energy limit where $ka_{2\text{D}}\ll1$, the atom-dimer
$s$- and $p$-wave amplitudes may be expanded as
\begin{equation}
f_s(k)\approx \frac{2\pi}{ \ln[1/(ka_\text{ad})]+i\pi/2},
\hspace{8mm}
f_p(k)\approx 4 s_\text{ad}k^2,
\label{eq:asymp}
\end{equation}
with the atom-dimer scattering length and surface, $a_\text{ad}$ and
$s_\text{ad}$, describing the $s$- and $p$-wave low-energy behavior,
respectively. Matching in the regime $ka_{2\text{D}}\ll1$, we find
\begin{equation} \notag
\frac{a_\text{ad}}{a_{2\text{D}}}=\left\{\begin{array}{r} 1.26\\ 2.29 \\
    1.02 \end{array}\right.
\hspace{-1mm},\hspace{1mm}
\frac{s_\text{ad}}{a_{2\text{D}}^2}=\left\{\begin{array}{rl} -2.92 &
   \hspace{4mm} m_\up/m_\down=1\\ -1.54 & \hspace{4mm} m_\up/m_\down=m_\text{K}/m_\text{Li}\\
    -0.44 & \hspace{4mm} m_\up/m_\down=m_\text{Li}/m_\text{K} \end{array}\right. \hspace{-2mm}
\end{equation}
{The low-energy quantities are shown as a function of mass ratio in
Fig.~\ref{fig:lowe}. We see that the scattering length increases
monotonically with $m_\up/m_\down$, whereas the scattering surface
displays a series of divergences, related to the appearance of trimers.}
The atom-dimer scattering length for equal masses has also been
estimated from a QMC calculation of the pairing gap in the BEC regime,
but they instead obtain
$a_\text{ad}\approx1.7a_{2\text{D}}$~\cite{Bertaina2011}.  This
disagreement with our exact few-body calculation is most likely
because the equation of state has a complicated dependence on
$a_\text{ad}$, making the determination of $a_\text{ad}$ difficult.
For large $m_\up/m_\down$, we find $a_\text{ad}\approx
0.845a_{2\text{D}} \ln(\sqrt2 e^{\gamma/2}m_\up/\mu)$, with the Euler
constant $\gamma\approx0.577$. A logarithmic dependence on mass ratio
was also found in the 3D case~\cite{PhysRevA.67.010703}.
{Additionally, from Eq.~(\ref{eq:asymp}) we calculate the
  asymptotic behavior of the cross section and this is shown in
  Fig.~\ref{fig:elastic}.} A surprising feature is that the $s$-wave
  scattering cross section is completely determined by $a_\text{ad}$
  for $m_\up\lesssim m_\down$, even though all coefficients in the
  low-energy expansion of the scattering amplitude would be expected
  to have a characteristic scale of $a_{2\text{D}}$. A similar result
  appears in 3D for $m_\up=m_\down$~\cite{Levinsen:2011uq}.

%


\section{Trimers}
\begin{figure}
\centering
\includegraphics[width=.95\linewidth]{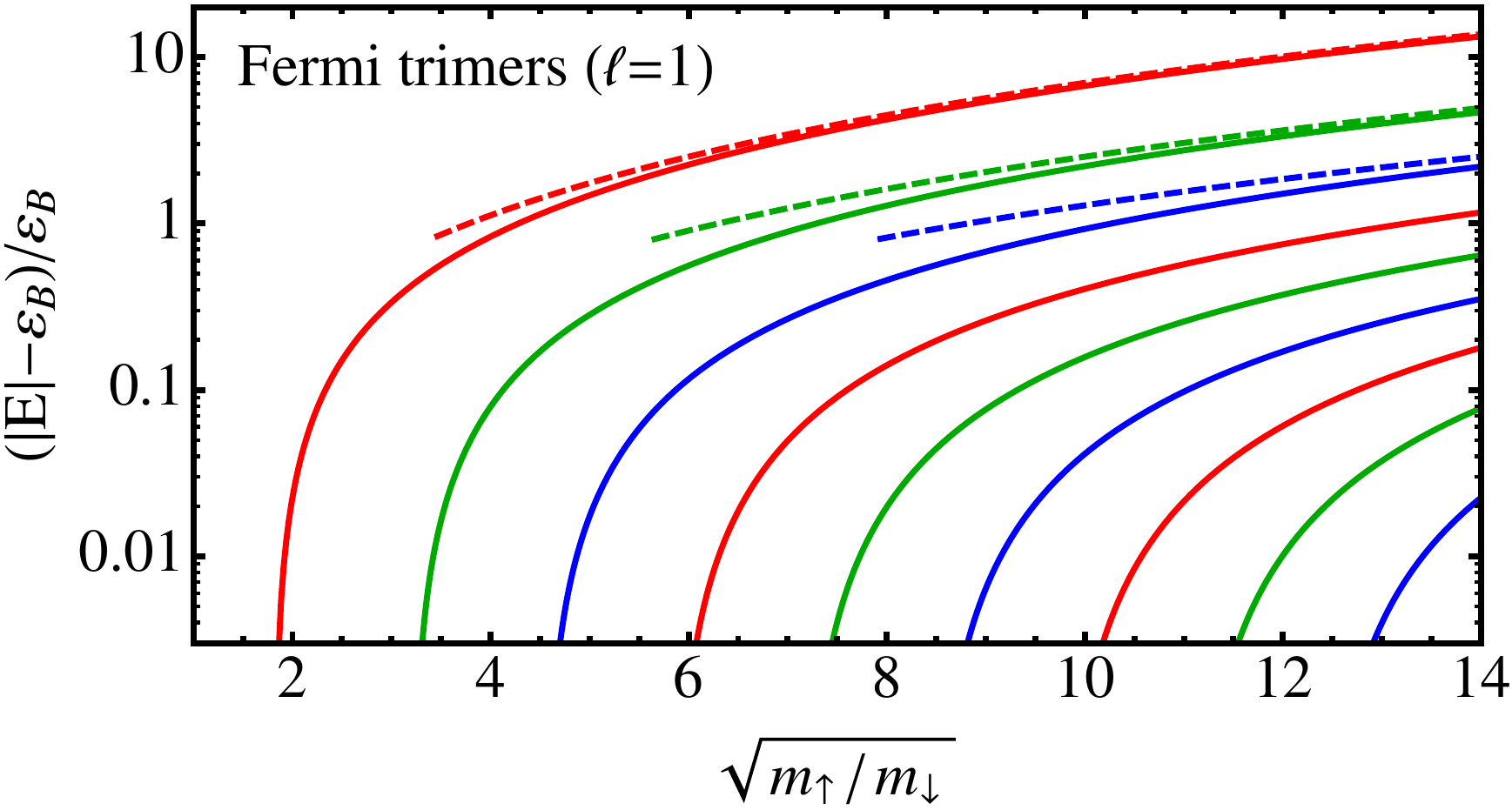}
\caption{(Color online) Trimer energies (solid lines) as a function of
mass ratio. Our results match those of
Ref.~\cite{PhysRevA.82.033625}. 
Dashed lines are the hydrogen-like spectrum, Eq.~(\ref{eq:hydrogen}).
\label{fig:BO}}
\end{figure}

We now turn to trimers, which formally appear as an energy pole in
Eq.~(\ref{eq:stm}) for odd partial waves. The spectrum of $p$-wave
trimers as a function of mass ratio is shown in Fig.~\ref{fig:BO}.
The appearance of trimers at large mass ratios in 2D may be elucidated
by applying the Born-Oppenheimer
approximation~\cite{BornOppenheimer},
which shows that the effective potential between heavy fermionic atoms
mediated by the light atom at distances $\lesssim a_{2\text{D}}$ is
similar to the electron potential in a hydrogen atom confined to
2D. To see this, assume that the state of the light atom at position
$\r$ adiabatically adjusts itself to the positions $\pm\R/2$ of the
heavy atoms.  Atom-dimer scattering in odd partial wave channels is
described by the symmetric light-atom
wavefunction~\cite{PhysRevA.67.010703}
\begin{equation}
\psi_\mathbf{R}(\mathbf r) \propto 
K_0(\kappa(R)|\mathbf{r-R/2}|) +
K_0(\kappa(R)|\mathbf{r+R/2}|),
\label{eq:wave}
\end{equation}
where the modified Bessel function of the second kind $K_0(\kappa r)$
is the decaying solution of the free Schr{\"o}dinger equation with
energy $\epsilon(R) = -\kappa(R)^2 / 2m_\downarrow$. The singularities
at the positions of the heavy atoms satisfy the Bethe-Peierls boundary
condition in 2D: For $\tilde\r=\r\pm\R/2$ this is $\left[\tilde
  r(\psi)'_{\tilde r}/\psi\right]_{\tilde \r\rightarrow\0} =
1/\ln(\tilde r e^\gamma/2a_{2\text{D}})$. Using the asymptotic form,
$K_0(x)\stackrel{x\to0}{=} -\ln(x e^\gamma/2)$, we then obtain the
condition $\ln\left(-\frac{\epsilon(R)}{\varepsilon_B}\right) =2
K_0\left( \sqrt{-\frac{\epsilon(R)}{\varepsilon_B}
  }\frac{R}{a_{2\text{D}}}\right).$

In the second stage of the Born-Oppenheimer approximation, the
Schr{\"o}dinger equation of the heavy particles is solved using
$\epsilon(R)$ as the effective interaction potential.  In the limit
$R\ll a_{2\text{D}}$, the potential becomes $\epsilon(R) \approx
-\frac{2 \varepsilon_B }{e^\gamma} \frac{a_{2\text{D}}}{R}$.  Thus,
the spectrum of the deepest bound trimer states is hydrogen-like and
given by the well-known result (appropriately shifted by the dimer
binding energy)~\cite{flugge1952}
\begin{equation}
E_n = -\frac{m_\uparrow}{e^{2\gamma} m_\downarrow}
\frac{\varepsilon_B}{2(n+1/2)^2} -\varepsilon_B,
\quad
\label{eq:hydrogen}
\end{equation}
with integer quantum number $n\geq\ell$. This also implies that deeply
bound trimers with different $\ell$ are degenerate, as was found in
Ref.~\cite{PhysRevA.82.033625}. Furthermore, we note that the
wavefunction (\ref{eq:wave}), and thus the above arguments, apply
equally well to heteronuclear bosons in even partial waves.

A similar scenario for large mass ratios was recently
predicted~\cite{PhysRevA.84.062704} in the context of a 3D Fermi gas
close to a narrow interspecies Feshbach resonance, characterized by a
large effective range $R^*$.  The effective interaction between the
two heavy fermions mediated by the light atom goes like $-1/R^{2}$ in
the range $R^*\ll R\ll a_s$, while at shorter ranges, $R\ll R^*$, this
behavior is replaced by an attractive $1/R$
potential~\cite{petrov-fewatom}, leading to a crossover from an
Efimovian spectrum for the weakly bound trimers to a hydrogen-like
spectrum for the deepest trimers~\cite{PhysRevA.84.062704}.

Referring to Fig.~\ref{fig:BO}, for the deepest bound states, we find
very good agreement with the hydrogen-like spectrum
(\ref{eq:hydrogen}).  At a given mass ratio, the potential only
supports a finite number of bound states proportional to the number of
nodes of the heavy-atom wavefunction that fit in the hydrogen-like
part of the potential, $R\lesssim a_{2\text{D}}$. We estimate this
number by noting that in this regime the wavefunction of heavy atoms
is proportional to the Bessel function
$J_{2\ell}(2\sqrt{e^{-\gamma}(m_\up/m_\down)R/a_{2\text{D}}})$. Since
the wavefunction acquires an additional node each time the argument
increases by $\pi$, the number of trimers is proportional to
$\sqrt{m_\up/m_\down}$. This feature is clearly observed in
Fig.~\ref{fig:BO}.


\section{Three-body recombination}
Turning now to inelastic scattering, three-body recombination is the
process whereby two atoms bind into a shallow dimer with the released
energy carried away by a third atom.  To extract the recombination
rate from the atom-dimer scattering amplitude, we consider the
recombination process in reverse. This is illustrated in
Fig.~\ref{fig:feynman}(b) which shows how the recombination process is
described by the same diagram as inelastic atom-dimer scattering -- a
process in which no dimer remains after the scattering. The total
recombination rate is therefore proportional to the atom-dimer
inelastic scattering cross section~\cite{PhysRevA.78.043605}
$K(E) = v_\text{ad}\frac{\Phi_\text{ad}}{\Phi_\text{aaa}}
\sum_\ell\sigma_\ell^\text{inel}(E),$ 
where the atom-dimer relative speed is $v_\text{ad} = k/\mu_3$ and the
sum is over all partial waves. The phase space of an atom-dimer pair
$\Phi_\text{ad}$ and three atoms $\Phi_\text{aaa}$ are given in the
appendix.  The inelastic scattering cross section
$\sigma_\ell^\text{inel}$ is obtained by subtracting the elastic from
the total cross section, which is related to the imaginary part of the
scattering amplitude through the optical
theorem~\cite{LL3:1977,1986AmJPh..54..362A}
\begin{equation}
  \sigma^\text{tot}_\ell(E) =-\frac{1}{k} \Im[f_\ell(k)]
  (2-\delta_{\ell,0}).
\end{equation}
Thus we arrive at the recombination rate
\begin{equation}
K(E) = \pi\rho!
\frac{2m_\uparrow+m_\downarrow}{m_\uparrow^2m_\downarrow}
\frac{2k}{E}
\sum_\ell \sigma_\ell^\text{inel}(E),
\end{equation}
with the degeneracy factor $\rho = 2$ if $\uparrow\neq\downarrow$,
\emph{i.e.}  for fermions or heteronuclear bosons, and $\rho=3$ for
identical bosons.

\begin{figure}
\centering
\includegraphics[width=.95\linewidth]{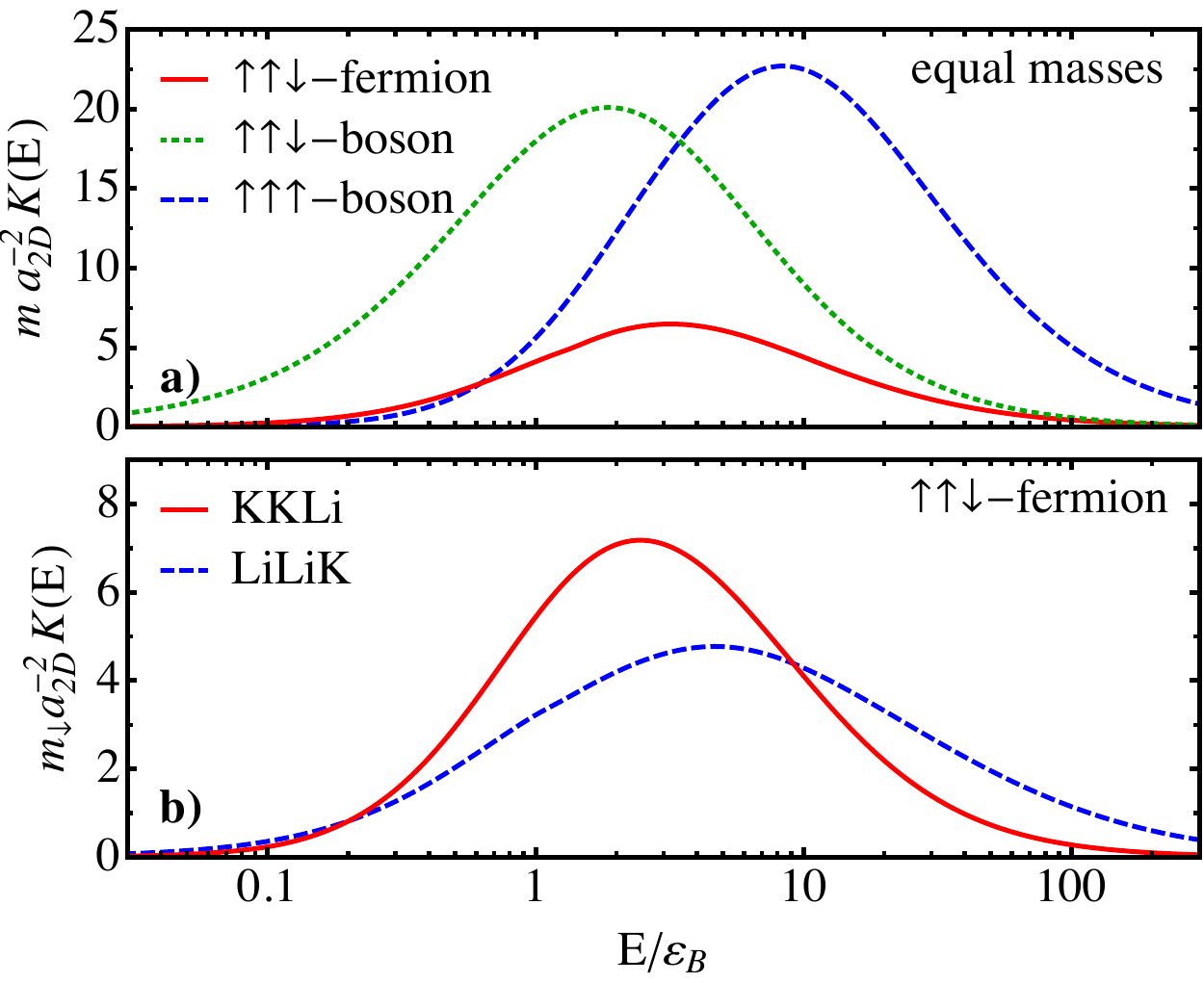}
\caption{(Color online) The recombination rate as a function of energy for (a)
  $m_\up=m_\down$ and (b) the Li-K heteronuclear mixture.
In the low energy limit $E/\eb \ll 1$,
the result for identical bosons matches the purely $s$-wave calculation 
in Ref.~\cite{PhysRevA.83.052703}.
\label{fig:K}}
\end{figure}

{ 

Figure~\ref{fig:K} shows the energy-dependent recombination rate for
atoms of equal masses and for the Li-K fermionic mixture. The rate
is expected to be maximal for scattering at energies of the order of
the energy of the bound pair in the final state, $\eb$, and indeed
this is seen to be the case.  Moreover, we find that several partial
waves are of the same order of magnitude in this
regime. Surprisingly, we find that the $d$-wave channel dominates
recombination of K-K-Li fermionic atoms and of three identical
bosons for most of the plotted energy range, whereas the $p$-wave
channel is dominant for equal mass and Li-Li-K fermions, as well as
for heteronuclear bosons.  Similar results were found for three
identical bosons in 3D~\cite{PhysRevA.78.043605}.

On the other hand, in the low energy limit of $E\ll\eb$ we find that
the fermionic rate is comparable to that of identical bosons. This is
manifestly different from the behavior in 3D, where the low-energy
recombination rate approaches a constant in the three-boson
problem~\cite{PhysRevLett.93.143201} while it is suppressed in Fermi
systems due to Pauli blocking with $K(E\rightarrow0)\sim
E$~\cite{PhysRevA.67.010703}. The low-energy behavior of the 2D
recombination rate originates from the logarithmic form of the
scattering amplitude, see Eq.~(\ref{eq:f}).
In all cases we find that $K(E)$ approaches zero faster than $E$,
which is a purely two-dimensional feature.  }

\begin{figure}
\centering
\includegraphics[width=.95\linewidth]{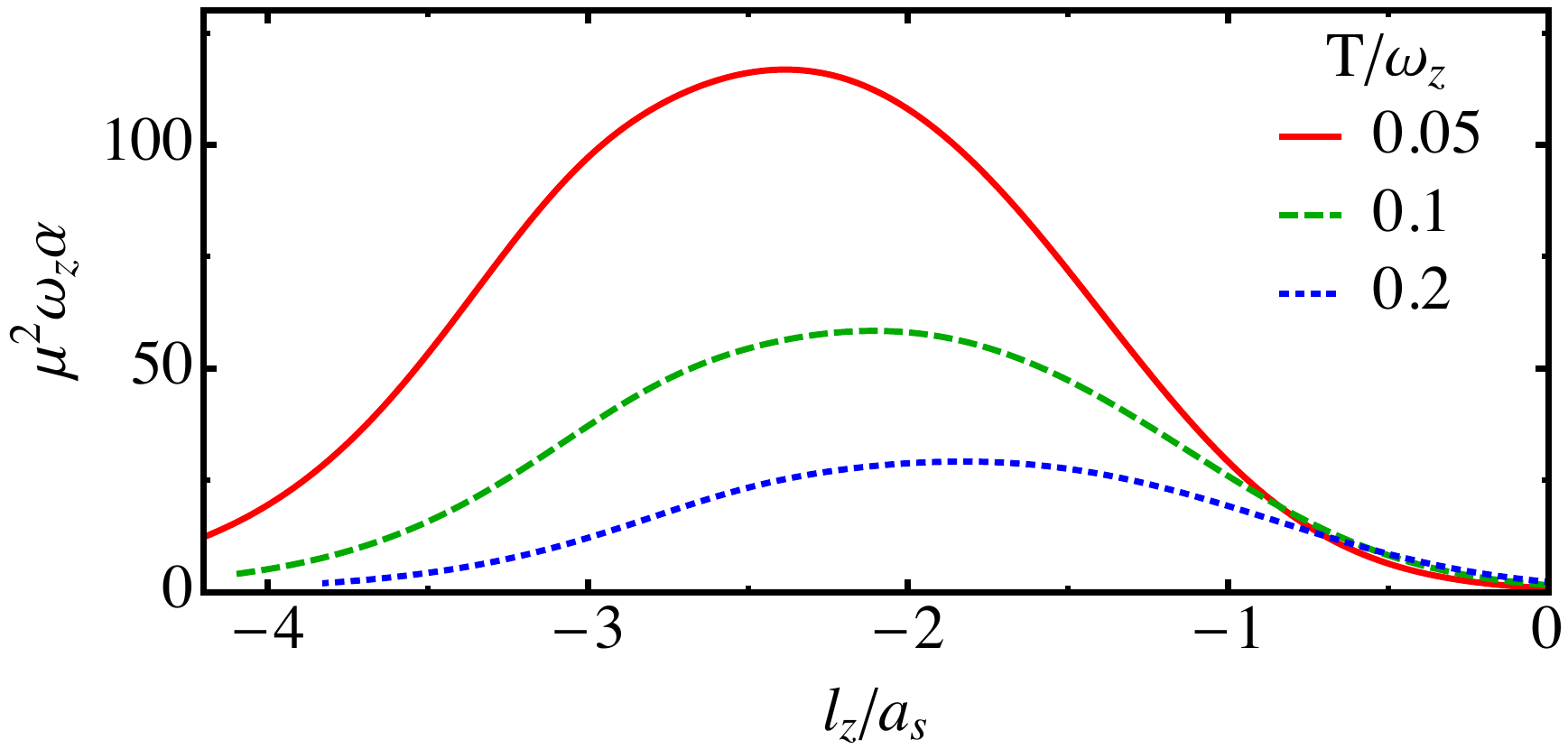}
\caption{(Color online) The event rate constant $\alpha$ 
  for fermions with $m_\up=m_\down$ as a function of (inverse) 3D scattering length.
  \label{fig:alpha}}
\end{figure}
%


To obtain an experimentally accessible quantity, we average $K(E)$
over energy with an appropriate thermal distribution.  This yields the
event rate constant per unit volume $\alpha$, defined such that the
number of recombination events per unit volume per unit time is
$\alpha n_\uparrow^2n_\downarrow$ for the process $\uparrow +\uparrow
+\downarrow \rightarrow \uparrow +\uparrow\downarrow$ in a
two-component Fermi mixture.  When the kinetic energy of the resulting
particles is greater than the height of the trap, the loss rate is
given by
$\dot n_\uparrow = 
-2\alpha_{\uparrow\uparrow\downarrow} n_\uparrow^2n_{\downarrow} 
-\alpha_{\downarrow\downarrow\uparrow} n_\downarrow^2n_\uparrow$. 
At sufficiently high temperatures, we employ the Boltzmann
distribution which gives~\cite{PhysRevA.78.030701,PhysRevA.83.052703}
\begin{equation}
\alpha(T) = 
\frac{\int_0^\infty dE\, E\,e^{-E/k_BT}K(E)}
{\rho!\int_0^\infty dE\, E\,e^{-E/k_BT}}.
\end{equation}
This quantity is illustrated in Fig.~\ref{fig:alpha} for a range of
typical temperatures. 
{
Similar to the recombination rate, the event rate constant is
expected to peak at a temperature of the order of the binding
energy.  Consequently, for increasing temperatures the peak value
decreases as $\alpha_\text{peak}\sim T^{-1}$ and the peak position,
$[l_z/a_s]_\text{peak}\sim -\ln(T)$, shifts from the BCS side of the
3D resonance towards unitarity.  Note that as the binding energy
decreases with decreasing $l_z/a_s$, the low-energy (high-energy)
recombination occurs to the right (left) of the peak in
Fig.~\ref{fig:alpha}. Towards the 3D resonance we
expect corrections to our results due to deviations from the 2D
limit, however these are small for the experimentally relevant
parameters considered in Fig.~\ref{fig:alpha}~\cite{note2}.  }


\section{Conclusion}
At present, heteronuclear fermionic Li-K
mixtures~\cite{PhysRevLett.100.053201,2011EL.....9633001R} are  
promising candidates for the observation of resonantly enhanced
atom-dimer scattering and trimer formation. 
Avenues towards observing the hydrogen-like spectrum predicted at
large mass ratio include using the recently predicted Feshbach
resonance in fermionic Li-Yb mixtures~\cite{PhysRevLett.108.043201}
as well as the application of a species selective optical lattice to enhance
the effective mass ratio~\cite{PhysRevLett.99.130407}.

We also expect our results to be important for ongoing experiments on
quasi-2D Fermi gases. In particular, three-body recombination will
limit the stability of repulsive Fermi gases and potentially generate
three-body correlations, thus impacting any experiment seeking to
realize itinerant ferromagnetism. Moreover, it would be interesting to
investigate how three-body losses might evolve into two-body losses in
the presence of a Fermi sea~\cite{Pietila2012,Ngampruetikorn2012}.

\acknowledgments
  We thank D.~S.~Petrov, S.~K.~Baur, M.~K\"ohl, and P.~Massignan for fruitful
  discussions. JL acknowledges support from the Carlsberg Foundation
  and from a Marie Curie Intra European grant within the 7th European
  Community Frame work Programme.
  MMP acknowledges support from the EPSRC under Grant No.\ EP/H00369X/2

\section{Appendix}
The $N$-particle phase space integral is defined such that
\begin{align}
\nonumber
\Phi&=\\
&\eta \int\left(\prod_{i=1}^N \frac{d^2\p_i}{(2\pi)^2}\right)
(2\pi)^2 \delta^{(2)}\left(\sum_{i=1}^N\p_i\right)
\delta\left(E-\sum_{i=1}^N\frac{\p_i}{2m_i}\right),
\end{align}
where indistinguishability of 
particles gives rise to the factor $\eta \le 1$. Using this definition, the atom-dimer 
and three-atom phase space is given by
\begin{align}
\Phi_\text{ad} = \mu_3/2\pi, \hspace{1.1cm}
\Phi_\text{aaa} = \eta \frac{E}{4\pi^2}
\frac{m_1m_2m_3}{m_1+m_2+m_3}.
\end{align}
For three particles, $\eta = 1/\rho!$ where $\rho$ is the number of 
identical particles. 
The phase space factors were derived in the bosonic case for equal masses in Ref.~\cite{PhysRevA.83.052703}.

\bibliography{Ref_Recomb}

\begin{thebibliography}{10}
\bibitem{note1}
The degeneracy comes from the quantization of angular momentum along the
  $z$-axis. In 2D, the degeneracy is $1$ if $\ell=0$ and $2$ otherwise, in
  contrast to $2\ell+1$ in 3D.
\end{thebibliography}
\bibliographystyle{eplbib}

\end{document}